\documentclass[twocolumn,showpacs,groupedaddress]{revtex4}
\usepackage{graphicx}

\begin{document}

\title{Atom-molecule conversion system subject to phase noises}
\author{H. Z. Shen$^{1}$, Xiao-Ming Xiu$^{1,2}$,  and X. X. Yi$^1$}
\affiliation{$^1$School of Physics and Optoelectronic Technology\\
Dalian University of Technology, Dalian 116024 China\\
$^2$Department of Physics, College of Mathematics and\\
Physics, Bohai University, Jinzhou
 121013, China}
\date{\today}

\begin{abstract}
The dynamics of  atom-molecule conversion system   subject to
dephasing  noises is studied in this paper. With the dephasing
master equation and the mean-field theory, we drive a Bloch equation
for the system, this equation is compared with the Bloch equation
derived by the Bogoliubov-Born-Green-Kirkwood-Yvon (BBGKY) hierarchy
truncation approach. Fixed points of the system are calculated by
solving both the Bloch equations and the master equation, comparison
between these two calculations  suggests that while in a short time
the mean-field theory is a good approximation for the atom-molecule
conversion system, a high order hierarchy truncation approach is
necessary  for the system in a long time scale. Although the MFT can
not predict correctly the fixed points, its prediction on the
stability of the fixed points are consistent with the BBGKY theory
for a wide range of parameters.
\end{abstract}

\pacs{03.75.-b, 03.75.Hh,  03.75.Gg} \maketitle

\section{Introduction}
In the realm of ultracold atom-molecule physics, association of
ultracold atoms into diatomic molecules  is an attractive subject.
It inspires many interests due to  its applications ranging from the
production of molecule Bose-Einstein condensates (BECs) to the study
of chemical reaction  and permanent electric dipole moments
\cite{Greiner2003,Heinzen2000,Zwierlein2003,
Regal2004,Junker2008,Jing2009,Qian2010,Inouye1998}.  Coherent
oscillations  between an atomic BEC and a molecular BEC have been
theoretically predicted\cite {Timmermans1999prl2691,Javanainen1999}
by the use of Gross-Pitaevskii (GP) equations
\cite{Vardi2001pra1,Santos2006,Li2009,Liu2010,Santos2010,Fu2010},
the results suggest  that the mean-field theory is a good formalism
to describe the conversion of atoms to molecules in the absence of
noise \cite{Timmermans1999prl199,Kohler2006}.

The noise may come from the inelastic collisions between the atoms
in the condensates and that in the non-condensate atoms, local
fluctuations and non-local fluctuations.  The noise may also come
from the random variation of the atom-molecule detuning or magnetic
field fluctuations in the Feshbach-resonance
setup\cite{Khripkov2011,Timmermans1999prl199,Duine2004,Brouard2005}.
\textbf{The presence of noise can dephase the Bose-Einstein
condensates and strongly limit the validity of the
Gross-Pitaevskii(GP) equations. There have been several theoretical
studied going beyond the GP equations, for example, based on the
time-dependent field theory, the dynamics of the atom-molecule
conversion system was studied in \cite{Javanainen2002,Naidon2008},
where the noise comes from nonlocal fluctuations due to the
time-dependent pair correlations, and within the two-model
approximation, the authors in Ref. \cite{Khripkov2011,Cui2012}
explored the master equation to investigate the atom-molecule
conversion system.}

Earlier study on a bimodal decoherence-free condensate show that the
mean-field theory(MFT) may fail near the dynamical
instability\cite{Anglin2001,Vardi2001prl568}, this inspires us to
explore whether the MFT is valid for the atom-molecule conversion
system with noise (dissipation and dephasing). The effect of
dissipation on the dynamics of the atom-molecule conversion system
was studied in \cite{Cui2012}. In this paper we will focus on the
effect of dephasing \textbf{within the two-mode approximation}. We
show that the dynamics of the dephasing atom-molecule conversion
system is well treated by the MFT in a short time scale, but it
fails to give a correct prediction about the system at a long time
scale. This suggests us to use the high order of   BBGKY hierarchy
truncation\cite {Anglin2001,Vardi2001prl568} to explore the
atom-molecule conversion system subject to dephasing noises.

The remainder of the paper is organized as follows. In Sec. {\rm
II}, we introduce the dephasing  master equation and derive a Bloch
equation for the system, the solution of the Bloch equation without
dephasing is presented and discussed. In Sec. {\rm III}, we
calculate the fixed points of the system with the MFT and compare
these fixed points with that by  analytically solving the master
equation. The Bloch equation derived from  the BBGKY hierarchy
equation is presented  in Sec. {\rm IV}.  In Sec. {\rm V}, we
discuss the stability and the feature of the fixed points from both
the MFT and the BBGKY hierarchy truncation. Discussion and
conclusions are given in Sec. {\rm VI}.
\section{Model}

We consider the simplest model for the atom-molecule conversion
system. By the two-mode approximation, the model Hamiltonian can be
written as\cite{Khripkov2011,Vardi2001pra1,Kostrun2000}
\begin{eqnarray}
\hat{H}=\frac{\varepsilon }{2}\hat{a}^{\dagger }\hat{a}+\frac{g}{2}(\hat{a}%
^{\dagger }\hat{a}^{\dagger }\hat{b}+\hat{b}^{\dagger }\hat{a}\hat{a}),
\label{Ham1}
\end{eqnarray}%
where $\hat{a}$ and $\hat{b}$ represent annihilation operators for
atom and molecule, respectively, $g$ denotes the strength of the
atom-molecule conversion, and $\varepsilon $ is the atomic binding
energy.

The master equation taking only the dephasing noise  into account
may be written into the following form\cite{Khripkov2011,Anglin1997}
\begin{eqnarray}
\dot \rho  =  - i[\hat H,\rho ] - \Gamma [\hat \ell ,[\hat \ell ,\rho ]],  \label{master2}
\end{eqnarray}%
where $\hat{\rho}$ is the density matrix of  system, $\Gamma $ is
the dephasing rate, the Lindblad operator  $\hat{\ell}$ is the
population difference,
\begin{eqnarray}
\hat{\ell}=2\hat{b}^{\dagger }\hat{b}-\hat{a}^{\dagger }\hat{a}.
\label{numberd3}
\end{eqnarray}
The total atom number operator $\hat{N}=2\hat{b}^{\dagger
}\hat{b}+\hat{a} ^{\dagger }\hat{a}$ is conserved since
$\frac{\partial \left\langle \hat{N} \right\rangle }{\partial t}=0,$
so the total atom number  $N$ is a constant that does not change
with time in the dynamics. Define,
\begin{eqnarray}
{{\hat L}_x} &=& \sqrt 2 \frac{{{{\hat a}^\dag }{{\hat a}^\dag }
\hat b + {{\hat b}^\dag }\hat a\hat a}}{{{N^{3/2}}}},\nonumber\\
{{\hat L}_y} &=& \sqrt 2 i\frac{{{{\hat a}^\dag }{{\hat a}^\dag }
\hat b - {{\hat b}^\dag }\hat a\hat a}}{{{N^{3/2}}}},\nonumber\\
{{\hat L}_z} &=& \frac{{2{{\hat b}^\dag }\hat b - {{\hat a}^\dag
}\hat a}}{N}, \label{angelmoment4}
\end{eqnarray}
where $\hat{L}_{z}$ denotes the  number difference between the atoms
and the molecules in the system,  $\hat{L}_{x}$ and $\hat{L}_{y}$
can be used to characterize the  coherence of atom-molecule
conversion. It is easy to prove that,
\begin{eqnarray}
\left[ \hat{L}_z,\hat{L}_x\right] &=& \frac{4i}{N}\hat{L}_y,\nonumber\\
\left[\hat{L}_z,\hat{L}_y \right] &=&  - \frac{4i}{N}\hat{L}_x,\nonumber\\
\left[\hat{L}_x,\hat{L}_y \right] &=& \frac{i}{N}(1 - \hat{L}_z) (1
+ 3\hat{L}_z) + \frac{4i}{N^2}. \label{commute5}
\end{eqnarray}
Notice that $\hat{L}_{x},\hat{L}_{y},\hat{L}_{z}$ are not the SU(2)
generators, because their commutation relations contain quadratic
terms of $\hat{L}_{z}.$ Nevertheless, in the small atom-molecule
number difference and large $N$ limit ($N\rightarrow \infty$),
$\hat{L}_{x}$, $\hat{L}_{y}$ and $\hat{L}_{z}$ really form a  sphere
since they satisfy,
\begin{eqnarray}
{({{\hat L}_x})^2} + {({{\hat L}_y})^2} &=&
\frac{1}{2}(1 + {{\hat L}_z}){(1 - \hat L{\ _z})^2} \nonumber\\
&+& \frac{2}{N}(1 - {{\hat L}_z}) + \frac{4}{{{N^2}}}{{\hat L}_z}.
\label{Bloch6}
\end{eqnarray}
We will call this sphere the generalized Bloch sphere even when the
system is far from the limits.
With these definitions, the Hamiltonian becomes
$\hat{H}=-\frac{\varepsilon }{4}N\hat{L}_{z}+\frac{g}{2%
\sqrt{2}}N^{3/2}\hat{L}_{x},$  and the master equation can be
rewritten as
\begin{eqnarray}
\dot \rho  =  - i[\hat H,\rho ] -   \Gamma {N^2}[{\hat L_z},[{\hat
L_z},\rho ]]. \label{masterlz7}
\end{eqnarray}

From this master  equation, the expectation values defined by  $F_{i}=$ $%
\langle {\hat L_i}\rangle  = Tr(\rho {\hat L_i}),i = x,y,z$ follow,
\begin{eqnarray}
\frac{{\partial {F_x}}}{{\partial t}} &=& \varepsilon {F_y} - 16\Gamma {F_x},\nonumber\\
\frac{{\partial {F_y}}}{{\partial t}} &=& - \varepsilon {F_x} -
\Delta {F_z} +
\frac{3}{2}\Delta \langle \hat L_z^2\rangle  - 16\Gamma {F_y} - R,\nonumber\\
\frac{{\partial {F_z}}}{{\partial t}} &=& 2\Delta {F_y},
\label{differgeneral8}
\end{eqnarray}
where $\Delta =g\sqrt{\frac{N}{2}},$ and $ R=\frac{1}{2}\Delta
+\frac{2\Delta }{N}.$   The lowest-order truncation of
Eq.~(\ref{differgeneral8}) is acquired by
approximating the second-order expectation values $\langle \hat{L}_{i}\hat{L}%
_{j}\rangle $ as products of the first-order expectations  $\langle \hat{L}%
_{i}\rangle $ and\ $\langle \hat{L}_{j}\rangle $ \cite{Anglin2001},
namely,
\begin{eqnarray}
\langle \hat{L}_{i}\hat{L}_{j}\rangle \approx \langle
\hat{L}_{i}\rangle \langle \hat{L}_{j}\rangle,  \label{MFT9}
\end{eqnarray}%
with this approximation, Eq.~(\ref{differgeneral8}) reduces to,
\begin{eqnarray}
\frac{{\partial {F_x}}}{{\partial t}} &=& \varepsilon {F_y} - 16\Gamma {F_x},\nonumber\\
\frac{{\partial {F_y}}}{{\partial t}} &=&  - \varepsilon {F_x}
- \Delta {F_z} + \frac{3}{2}\Delta F_z^2 - 16\Gamma {F_y} - R,\nonumber\\
\frac{{\partial {F_z}}}{{\partial t}} &=& 2\Delta {F_y}.
\label{differMFT10}
\end{eqnarray}
Next we discuss the situation with zero dephasing rate, $\Gamma =0$,
Eq.~(\ref{differMFT10}) follows,
\begin{eqnarray}
\frac{{\partial {F_x}}}{{\partial t}} &=& \varepsilon {F_y},\nonumber\\
\frac{{\partial {F_y}}}{{\partial t}} &=&  - \varepsilon {F_x}
 - \Delta {F_z} + \frac{3}{2}\Delta F_z^2 - R,\nonumber\\
\frac{{\partial {F_z}}}{{\partial t}} &=& 2\Delta {F_y}.
\label{differg=011}
\end{eqnarray}
Define $a=\varepsilon ^{2}+g^{2}N,$ $b=-\frac{3}{2}g^{2}N,$ and
$c=\frac{g^{2}N}{2}+2g^{2}-\varepsilon ^{2}F_{z0}$ with $F_{z0}$ the
initial value of $F_{z}$, the solution of Eq.~(\ref{differg=011})
can be obtained by solving,
\begin{equation}
\frac{\partial ^{2}F_{z}}{\partial
^{2}t}+aF_{z}+bF_{z}^{2}+c=0.\label{Fze}
\end{equation}
We notice that $b$ must not be zero here, otherwise $g=0$, which
would result in $\Delta = 0$ leading to  ${\dot F_z(t)} = 0,$ then
$F(t)_z \equiv  F_{z0}$, i.e., the state of system remains
unchanged. The solution of Eq. (\ref{Fze}) is,
\begin{eqnarray}
F_{z}=u_{2}-(u_{2}-u_{3})cn^{2}(k(t-t_{0}),m)-\frac{a-A}{2b},
\label{FZ}
\end{eqnarray}
where $cn(k(t-t_{0}),m)$ is the Jacobi elliptic cosine function.
$u_{1}>u_{2}>u_{3}$ and $u_{1}=n\cos \theta-\frac{A}{{2B}},$
$u_{2}=n\cos (\theta +\frac{4\pi }{3} )-\frac{A}{{2B}},$
$u_{3}=n\cos (\theta +\frac{2\pi }{3})-\frac{A}{{2B}},$ $n =
\frac{{A}}{{B}},$ $\cos(3\theta )=- \frac{1}{2}\left( {d{{\left(
{\frac{{2B}}{A}} \right)}^3} + 2} \right),$ $A=\sqrt{a^{2}-4bc},$
$B=b,$  $d =  - u_0^3 - \frac{{3A}}{{2B}}u_0^2$, and ${u_0} =
{F_{z0}} + \frac{{a - A}}{{2b}}$.

$F_z(t)$ is a periodic function of time with period
$T=\frac{2K(m)}{k}$, $k = \sqrt {\frac{{ - B({u_1} - {u_3})}}{6}}$,
and $K(m)= \int_0^{\frac{\pi }{2}} {\frac{1}{{\sqrt {1 - {m^2}{{\sin
}^2}(\varphi )} }}} d\varphi$ being  the first kind Legendre's
complete elliptic integral.  ${t_0}$ denotes the time when $F_z$
takes $F_{z0}$ that can be determined by solving Eq.~(\ref{FZ}).
Eq.~ (\ref{differg=011}) describes a rotation of the Bloch vector
$\mathbf{F}$, obviously the norm $\left\vert \mathbf{ F}\right\vert
$ is conserved in the MFT when the dephasing rate is zero.

\begin{figure}[h]
\centering
\includegraphics[angle=0,width=0.45\textwidth]{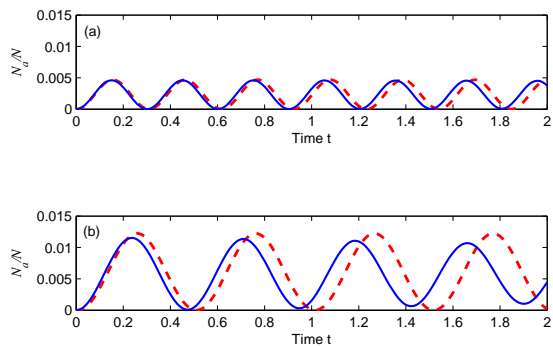}
\caption{(Color online) The number of atoms  in the atomic mode as a
function of time. The red-dashed line  represents Eq. (\ref{FZ}),
which is an analytical solution of the Bloch  equation with $\Gamma
= 0$ based on the mean-field theory. In contrast,   the numerical
simulation of the Liouville equation~(\ref{master2})  with  $\Gamma
= 0$ is shown by the blue-solid line with  ${F_{z0}} = 1, {F_{x0}} =
{F_{y0}} = 0$ at time $t=0$. Here and hereafter,
$\protect\varepsilon$ and $\Gamma$ are rescaled in units of $g$, and
$t$ is then in units of $1/g$. Hence all parameters are of
dimensionless. $N = 100$. (a) and (b) are for different
$\protect\varepsilon$. (a) $\protect\varepsilon = 25,$ and (b)
$\protect \varepsilon = 19.$} \label{analy:}
\end{figure}

In Fig.~\ref{analy:}, we plot the ratio of $N_a$ to $N$  as a
function of time. Two results are presented, one comes from Eq.
(\ref{FZ}), and another is obtained  by solving the master equation
with $\Gamma=0$ numerically.  We find that at a short time scale,
the two results are in good agreement, however, at a long time
scale, the two results are evidently different. This suggests that
the MFT is a good approximation to describe the dynamics of the
atom-molecule conversion system at a short time scale. Besides, the
binding energy of the atom can turn the system from self-trapping
regime (Fig~\ref{analy:} (a)) to tunneling regime (Fig~\ref{analy:}
(b)). This  can be understood as a conversion blockage due to the
energy difference  between  the atoms and \textbf{molecules. Note} that the
binding energy of the molecules is zero.

\section{Steady state and fixed points}

The fixed point of the system is defined by
\begin{eqnarray}
{{\dot F}_x} = {{\dot F}_y} = {{\dot F}_z} = 0.
\label{stablecondition14}
\end{eqnarray}
By this definition, we can obtain the fixed points in the MFT,
\begin{eqnarray}
{F_{xf}} = {F_{yf}} = 0,
{F_{zf}}=\frac{1}{3}\left( {1-\sqrt{1+3\left( {1+{4\mathord{\left/
{\vphantom {4 N}} \right. \kern-\nulldelimiterspace}N}}\right) }}\right).
\label{stablepoint15}
\end{eqnarray}
On the other hand, we can obtain the steady state by analytically
solving the master equation \textbf{Eq.~(\ref{master2}). Once} we have the
steady state of the system, the fixed points can be calculated by
the definition of $F_j.$ The steady state ${\rho _s}$ satisfies  the
following equation,
\begin{eqnarray}
{{\dot \rho }_s}=-i[\hat H,{\rho _s}]{\rm{ + }}\Gamma  \left( {2\hat
\ell {\rho _s}\hat \ell  - \hat \ell \hat \ell {\rho _s} - {\rho
_s}\hat \ell \hat \ell } \right) = 0. \label{stablestate}
\end{eqnarray}
It is easy to  prove that the off-diagonal elements of  density
matrix vanish in the steady state due to the dephasing. The proof is
as follows. Define Fock states  $\left| n \right\rangle \equiv
\left| {N - 2n,n} \right\rangle$ denoting  $(N-2n)$ atoms and $n$
molecules (${{n = 0, 1, 2,}} \cdots {{N/2}}$), we have the
following equation for the off-diagonal elements of the density
matrix,
\begin{eqnarray}
&&\frac{{\partial {\rho _{mn}}}}{{\partial t}} + i\left( {{a_m} -
{a_n}} \right){\rho _{mn}} \nonumber\\
&&+ 16\Gamma {\left( {m - n} \right)^2}{\rho _{mn}}
 + \xi \left( t \right) = 0,
\label{pmn}
\end{eqnarray}
where $\xi \left( t \right) = i({b_m}{\rho _{m - 1n}} + {c_m}{\rho
_{m + 1n}} - {b_n}{\rho _{mn - 1}} - {c_n}{\rho _{mn + 1}})$, ${a_n}
= \frac{\varepsilon }{2}(N - 2n)$, ${b_n} = \frac{g}{2}(\sqrt
{\left( {N - 2n + 1} \right) \left( {N - 2n + 2} \right)\left( n
\right)} )$, and ${c_n} = \frac{g}{2}(\sqrt {\left( {N - 2n}
\right)\left( {N - 2n - 1} \right)\left( {n + 1} \right)} )$. The
formal solution of  Eq.~(\ref{pmn}) is
\begin{widetext}
\begin{eqnarray}
{\rho _{mn}} = {e^{ - \left( {i\left( {{a_m} - {a_n}} \right) +
16\Gamma {{\left( {m - n} \right)}^2}} \right)t}}[\Xi  - \int \xi(t)
{e^{\left( {i\left( {{a_m} - {a_n}} \right) + 16\Gamma {{\left( {m -
n} \right)}^2}} \right)t}}dt], \label{solution of pmn}
\end{eqnarray}
\end{widetext}
where $\Xi $ is a constant determined by  the initial  condition  of
${{\rho _{mn}}}$. We find that when $t \to \infty$, ${\rho _{mn}}
\to 0 \left( {m \ne n} \right)$. This gives   the steady  state,
\begin{eqnarray}
{\rho _s} = \sum\limits_{n = 0}^{N/2} {{\rho _n}}  \left| n
\right\rangle \left\langle n \right|. \label{ps}
\end{eqnarray}
For the steady state, it is  required that  $[\hat H,{\rho _s}] =
0$, from which we obtain ${\rho _j} = {\rho _{j - 1}}$. This
together with $Tr{\rho _s} = 1$, we obtain ${\rho _0} = {\rho _1} =
{\rho _2} = \cdots = {\rho _{N/2}} = \frac{1}{{N/2 + 1}}$.
Collecting all together, we have,
\begin{eqnarray}
{\rho _s} = \sum\limits_{n = 0}^{N/2}  {\left( {\frac{1}{{N/2 + 1}}}
\right)\left| n \right\rangle \left\langle n \right|,}
\label{stablestatefinaly}
\end{eqnarray}
The fixed points ${F_{is}}, \ (i=x,y,z) $ of the system can be given
by the steady  state Eq.~(\ref{stablestatefinaly}) as  ${{F_{zs}} =
Tr({\rho _s}{{\hat L}_z}) = \sum\limits_{n = 0}^{N/2} {{{\hat L}_z}}
(\frac{1}{{N/2 + 1}})\left| n \right\rangle \langle n| = 0}.$ In the
same way, ${F_{xs}} = {F_{ys}} = 0$. Namely, the fixed point given
by  solving the master equation is,
\begin{eqnarray}
{F_{xs}} = {F_{ys}} = {F_{zs}} = 0. \label{stable p}
\end{eqnarray}

It is easy to find that the fixed points given by the MFT and the
master equation are different. This indicates that the MFT is not a
good approach to describe the atom-molecule conversion system at a
long time scale. This stimulates us to use the BBGKY hierarchy
truncation\cite {Anglin2001,Vardi2001prl568} to study the system.

\section{The BBGKY hierarchy of equations of motion}
As aforementioned, the differential equation for  the Bloch vector
up to the first order is  not a good treatment  at a long time
scale. Thus high order  expectation values is required.  In this
section, we will  obtain an improved theory to the MFT using the
next order of the Bogoliubov-Born-Green-Kirkwood-Yvon (BBGKY)
hierarchy of equation of motion.

Writing $\langle \hat L_z^2\rangle$ in Eq. (\ref{differgeneral8}) in
terms of the  following expectation value,
\begin{eqnarray}
K_{ij}=\langle \hat{L}_{i}\hat{L}_{j}+\hat{L}_{j}\hat{L}_{i}\rangle
-2\langle \hat{L}_{i}\rangle \langle \hat{L}_{j}\rangle ,i,j=x,y,z,
\label{K18}
\end{eqnarray}
and truncating the   BBGKY hierarchy of  equations of motion for the
first- and second-order operators
$\hat{L}_{i},\hat{L}_{i}\hat{L}_{j}$
\cite{Anglin2001,Vardi2001prl568},
\begin{eqnarray}
\langle {{\hat L}_i}{{\hat L}_j}{{\hat L}_k}\rangle &\approx&
\langle {{\hat L}_i}{{\hat L}_j}\rangle \langle {{\hat L}_k}\rangle
+ \langle {{\hat L}_i}\rangle \langle {{\hat L}_j}{{\hat
L}_k}\rangle
\nonumber\\
&+& \langle {{\hat L}_i}{{\hat L}_k}\rangle \langle {{\hat
L}_j}\rangle
 - 2\langle {{\hat L}_i}\rangle \langle {{\hat L}_j}\rangle \langle {{\hat
 L}_k}\rangle,
\label{LLL19}
\end{eqnarray}
we get  the following set of  equations for the  first- and
second-order moments,
\begin{eqnarray}
\frac{{\partial {F_x}}}{{\partial t}} &=& \varepsilon {F_y} - 16\Gamma {F_x},\nonumber\\
\frac{{\partial {F_y}}}{{\partial t}} &=&  - \varepsilon {F_x} -
\Delta {F_z} + \frac{3}{2}\Delta \left( {\frac{1}{2}{K_{zz}} +
F_z^2} \right)- 16\Gamma {F_y} - R \nonumber\\
\frac{{\partial {F_z}}}{{\partial t}} &=& 2\Delta {F_y}, \nonumber\\
\frac{{\partial {K_{xx}}}}{{\partial t}} &=& 2\varepsilon {K_{xy}}
- 32\Gamma {K_{xx}} + 32\Gamma {K_{yy}} + 64\Gamma F_y^2, \nonumber\\
\frac{{\partial {K_{yy}}}}{{\partial t}} &=&  - 2\varepsilon
{K_{xy}}
- 2\Delta {K_{yz}} + 6\Delta {F_z}{K_{yz}} - 32\Gamma {K_{yy}}\nonumber\\
 &+& 32\Gamma {K_{xx}} + 64\Gamma F_x^2,\nonumber\\
\frac{{\partial {K_{zz}}}}{{\partial t}} &=& 4\Delta {K_{yz}},\nonumber\\
\frac{{\partial {K_{xy}}}}{{\partial t}} &=&  - \varepsilon {K_{xx}}
- \Delta {K_{xz}} + 3\Delta {F_z}{K_{xz}} + \varepsilon {K_{yy}} -
64\Gamma {K_{xy}}\nonumber\\
 &-& 64\Gamma {F_x}{F_y},\nonumber\\
\frac{{\partial {K_{yz}}}}{{\partial t}} &=& 2\Delta {K_{yy}} -
\varepsilon {K_{xz}} - \Delta {K_{zz}} + 3\Delta {F_z}{K_{zz}}
- 16\Gamma {K_{yz}}\nonumber\\
\frac{{\partial {K_{xz}}}}{{\partial t}} &=& 2\Delta {K_{xy}} +
\varepsilon {K_{yz}} - 16\Gamma {K_{xz}}.
 \label{Beyondequation20}
\end{eqnarray}
Eq.~(\ref{Beyondequation20}) was called Bogoliubov backreaction
equations\cite{Anglin2001,Vardi2001prl568} (BBR),  because the
fluctuations $K_{ij}$ are driven  by the mean-field Bloch vector
\textbf{F}, which is physically described by the Bogoliubov theory.
In turn, the Bloch vector is affected by the fluctuations
$K_{\text{ij}}$. This  backreaction makes the trajectory of the
system not confined to the surface of the generalized Bloch sphere,
which is a reminiscence  of  the effect of  \textbf{dephasing}.

\begin{figure}[h]
\centering
\includegraphics[angle=0,width=0.45\textwidth]{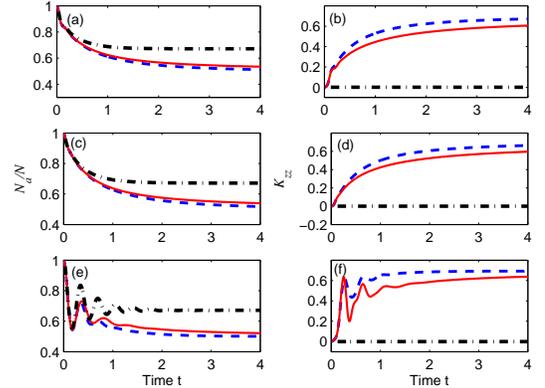}
\caption{(Color online) $N_a/N$ and $K_{zz}$ versus time. The
results are obtained by the mean-field theory (black-dash-dotted
line), Bogoliubov backreaction equations (red-solid line), and the
numerical solution of the master equation (blue-dashed line). The
initial condition of the system are the same as in
Eq.~(\ref{initialvalue16}). Parameters chosen are $g=1, N=100,
\protect\varepsilon =30, \Gamma =1$ for (a) and (b),
$\protect\varepsilon =40, \Gamma =1.8$ for (c) and (d), and
$\protect\varepsilon =10, \Gamma =0.2$ for (e) and (f).}
\label{beyMFT:}
\end{figure}

We plot the time evolution of the atom number ${N_a}$  and the
fluctuation ${K_{{{zz}}}}$ given by BBR  and MFT in
Fig.~\ref{beyMFT:}. The results from numerically solving the  master
equation  Eq.~(\ref{master2}) is also presented. To plot the figure,
the following initial condition
\begin{eqnarray}
\begin{array}{l}
{F_z} =  - 1,\\
{K_{xx}} = {K_{yy}} = 4\left( {N - 1} \right)/{N^2},\\
{F_x} = {F_y} = {K_{zz}} = {K_{xy}} = {K_{xz}} = {K_{yz}} = 0
\end{array}  \label{initialvalue16}
\end{eqnarray}
is taken, the corresponding quantum state is  the molecular vacuum
state $\left\vert N,0\right\rangle$.
We find that the results given by the BBR equations are in good
agreement with that by numerically solving the master equation. The
results by the MFT are different from those at a long time scale.
This difference comes from the fluctuations $K_{ij}$, which are
ignored in the MFT. Noticing the fixed points given by the BBR
equations are the same as that by the numerical method but different
from those by MFT, we emphasize that the stability of the fixed
points by MFT and BBR equations are the same for a wide range of
parameters in the space spanned by $F_x$, $F_y$ and $F_z$, this is
due to the linear coupling between the Bloch vector \textbf{F} and
the fluctuations $K_{ij}$ in \textbf{F}, see the first three
equations in Eq.(\ref{Beyondequation20}).

\begin{figure}[h]
\centering
\includegraphics[angle=0,width=0.45\textwidth]{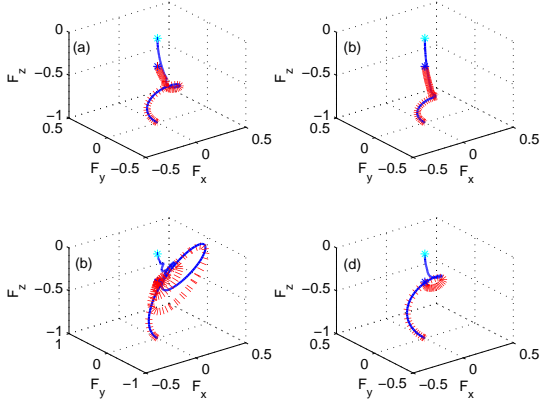}
\caption{(Color online) $F_z$ as a function of time by the
mean-field theory (red-dotted line) and Bogoliubov backreaction
equation (blue-solid line). The black-star denotes the fixed point
by the MFT in Eq.~(\ref{differMFT10}) and the green-star denotes the
fixed point by BBR in Eq.~(\ref{Beyondequation20}). The parameters
of (a), (b) and (c) are the same as in Fig.~\ref{beyMFT:}. In figure
(d), $g=1,\protect\varepsilon =15, \Gamma =0.8, N=100$.}
\label{transition:}
\end{figure}
\section{Stability of the fixed points with $\varepsilon=0$}
\begin{figure}[h]
\centering
\includegraphics[angle=0,width=0.45\textwidth]{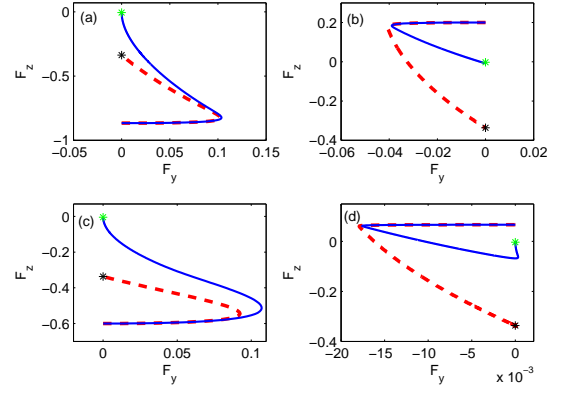}
\caption{(Color online) This plot shows the fixed points and how the
system approaches the fixed points. The black  and green stars
denote the location of the stable junction fixed point by the  MFT
 and by the  BBR (see, Eq.~(\ref{fixedpointBBR})),
respectively. The red-dashed line (MFT) and the blue-solid line(BBR)
show how the system approaches the fixed points. The initial state
of the system is   $\left| {{\psi _0}} \right\rangle = \left| n
\right\rangle $. Parameters chosen are $g=1, \Gamma=10, N=300$,
$\left| {{\psi _0}} \right\rangle  = \left| {10} \right\rangle $ for
(a), $\Gamma=12,$ $\left| {{\psi _0}} \right\rangle  = \left| {90}
\right\rangle $ for (b),  $ \Gamma=4$, $\left| {{\psi _0}}
\right\rangle  = \left| {30} \right\rangle $ for (c) and $
\Gamma=24$, $\left| {{\psi _0}} \right\rangle  = \left| {80}
\right\rangle $ for (d).} \label{junction:}
\end{figure}

\begin{figure}[h]
\centering
\includegraphics[angle=0,width=0.45\textwidth]{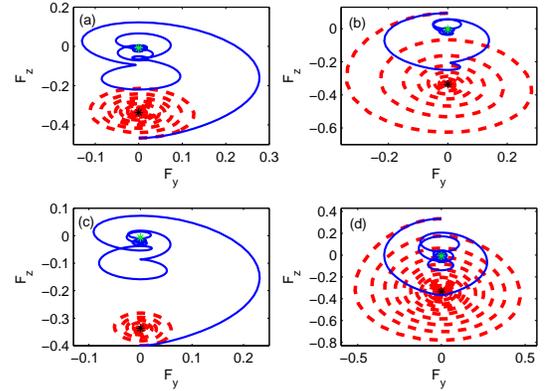}
\caption{(Color online) Black star and green star denote  the
location of the stable focus fixed points predicted by MFT
 and BBR, respectively. The red-dashed (MFT)
and the blue-solid (BBR) show  the trajectories for the system from
initial state to fixed points. Parameters chosen are $g=1,
\Gamma=0.12, N=300$, $\left| {{\psi _0}} \right\rangle  = \left|
{40} \right\rangle $ for (a), $\Gamma=0.2$, $\left| {{\psi _0}}
\right\rangle  = \left| {82} \right\rangle $ for (b), $\Gamma=0.16$,
$\left| {{\psi _0}} \right\rangle  = \left| {45} \right\rangle $ for
(c), and $ \Gamma=0.1$, $\left| {{\psi _0}} \right\rangle  = \left|
{100} \right\rangle $ for (d).} \label{focus:}
\end{figure}
In this section, we will discuss stability of the fixed points from
both the MFT and the BBGKY hierarchy. For the reason of simplicity,
let us consider the situation of zero atomic binding energy,
$\varepsilon=0$. In this case,  Eq.~(\ref{differMFT10}) reduces to,
\begin{eqnarray}
\begin{array}{l}
\frac{{\partial {F_y}}}{{\partial t}} =
- \Delta {F_z} + \frac{3}{2}\Delta F_z^2 - 16\Gamma {F_y} - R,\\
\frac{{\partial {F_z}}}{{\partial t}} = 2\Delta {F_y}.
\end{array}  \label{differe=012}
\end{eqnarray}
By the Jacobian matrix defined by
\begin{eqnarray}
J = {\left( {\begin{array}{*{20}{c}}
{\frac{{\partial P}}{{\partial {F_y}}}}&{\frac{{\partial P}}{{\partial {F_z}}}}\\
{\frac{{\partial Q}}{{\partial {F_y}}}}&{\frac{{\partial Q}}{{\partial {F_z}}}}
\end{array}} \right)_{\left( {{F_{xf}},{F_{yf}},{F_{zf}}} \right)}} ,  \label{J12}
\end{eqnarray}
we can study the stability of the fixed points in the MFT. Here
$P=-\Delta F_{z}+\frac{3 }{2}\Delta F_{z}^{2}-16\Gamma F_{y}-R$,
$Q=2\Delta F_{y}$. The eigenvalues of the Jacobi matrix $J$ would
determine the stability of the fixed points, which can be given by
simple calculations,  $\lambda_{\pm} = \frac{1}{2}( - 16\Gamma \pm
\sqrt {256{\Gamma ^2} - 4N{g^2}(1 - 3{F_{zf}})} ) = \frac{1}{2}( -
16\Gamma  \pm \sqrt {256{\Gamma ^2} - {\rm{4}}N{g^2}\sqrt {1{\rm{ +
}}3(1{\rm{ + }}4/N)} } )$. If it  is satisfied that
\begin{eqnarray}
64{\Gamma ^2} \ge N{g^2}\sqrt {1{\rm{ + }}3\left( {1{\rm{ + }}4/N} \right)},
\label{condition1}
\end{eqnarray}
Jacobi matrix $J$ has two negative roots, the fixed point
(\ref{stablepoint15})  is a stable junction fixed point. When
\begin{eqnarray}
64{\Gamma ^2} \le N{g^2}\sqrt {1{\rm{ + }}3\left( {1{\rm{ + }}4/N} \right)},
\label{condition2}
\end{eqnarray}
Jacobi matrix $J$ has two conjugate complex roots, in this case the
fixed point (\ref{stablepoint15}) is a stable  focus fixed point.

Now we turn our discussion to the fixed points given by the  BBGKY
hierarchy of equation of motion. To compare the stability of fixed
points  by the BBGKY with the prediction by the MFT, we restrict the
discussion in the space spanned by $F_x$, $F_y$ and $F_z$. This
means that the fluctuations which   drive the system away from the
fixed points (steady state) occur only in $F_x$, $F_y$ and $F_z$. We
start with the fixed points in the 9-dimensional space. By the same
definition as  in the MFT, we obtain the fixed point in the BBGKY
Eq.~(\ref{Beyondequation20}) ($\varepsilon = 0$)
\begin{eqnarray}
\begin{array}{l}
{F_{xB}} = {F_{yB}} = {F_{zB}} = {K_{xzB}} = {K_{xyB}} = {K_{yzB}} = 0,\\
{K_{zzB}} = \frac{2}{3} + \frac{8}{{3N}},\\
{K_{xxB}} = {K_{yyB}} = \frac{1}{2}{K_{zzB}}.
\end{array}
 \label{fixedpointBBR}
\end{eqnarray}%
As mentioned, we discuss the case where fluctuations are only in
$F_x$, $F_y$ and $F_z$. For $\varepsilon=0$, $F_x$ decouples with
$F_y$ and $F_z$, then the discussion reduce to discuss fluctuations
only in $F_y$ and $F_z$,
\begin{eqnarray}
\begin{array}{l}
{F_y} \to {F_{yB}} + \delta {f_y},\\
{F_z} \to {F_{zB}} + \delta {f_z},
\end{array}
\label{fluctuations}
\end{eqnarray}
Substituting Eq.~(\ref{fluctuations}) into
Eq.~(\ref{Beyondequation20}), we have,
\begin{eqnarray}
\frac{{\partial \delta {f_y}}}{{\partial t}} &=&  - \varepsilon
\delta {f_x} - \Delta \delta {f_z} + \frac{3}{2}\Delta
\left( {\frac{1}{2}{K_{zzB}} + 2{F_{zB}}\delta {f_z}} \right)\nonumber\\
&-& 16\Gamma \delta {f_y} - R,\nonumber\\
\frac{{\partial \delta {f_z}}}{{\partial t}} &=& 2\Delta \delta
{f_y}. \label{smallfuuctuations}
\end{eqnarray}
By the same discussion as in  Eq.~(\ref{J12}), the Jacobian matrix
in this case is,
\begin{eqnarray}
J' = {\left( {\begin{array}{*{20}{c}}
{\frac{{\partial P'}}{{\partial \delta {f_y}}}}&{\frac{{\partial P'}}
{{\partial \delta {f_z}}}}\\
{\frac{{\partial Q'}}{{\partial \delta {f_y}}}}&{\frac{{\partial Q'}}
{{\partial \delta {f_z}}}}
\end{array}} \right)_{\left( {{F_{xB}},{F_{yB}},{F_{zB}}} \right)}},
\label{JB}
\end{eqnarray}
where $$P' =  - \varepsilon \delta {f_x} - \Delta \delta \ {f_z} +
\frac{3}{2}\Delta \left( {\frac{1}{2}{K_{zzB}} + 2\ {F_{zB}}\delta
{f_z}} \right) $$$$- 16\Gamma \delta {f_y}
 - R,$$
and $Q' = 2\Delta \delta {f_y},$ the eigenvalues of the Jacobi
matrix $J'$ are $\lambda '_{\pm} = \frac{1}{2}( - 16\Gamma  \pm
\sqrt {256{\Gamma ^2} - 4N{g^2}(1 - 3{F_{zB}}} ) = \frac{1}{2}( -
16\Gamma  \pm \sqrt {256{\Gamma ^2} - 4N{g^2}} ).$ If
\begin{eqnarray}
64{\Gamma ^2} \ge N{g^2},
\label{condition3}
\end{eqnarray}
all $\lambda '_{\pm}$ are  negative, the fixed point
(\ref{fixedpointBBR}) is a stable junction fixed point. Otherwise if
\begin{eqnarray}
64{\Gamma ^2} \le N{g^2},
\label{condition4}
\end{eqnarray}
$\lambda '_{\pm}$ are complex  and their real parts are negative,
the fixed point (\ref{fixedpointBBR}) is then a stable focus fixed
point.

When the parameters satisfy simultaneously  Eq.~(\ref{condition1})
and Eq.~(\ref{condition3}), the stability of the fixed points are
the same in the MFT and the BBGKY, i.e., the fixed points are stable
junction point, see Fig.~\ref{junction:}.  In this situation, the
system \textbf{approaches} the fixed points straightforwardly.
When the parameters satisfy both Eq.~(\ref{condition2}) and
Eq.~(\ref{condition4}), the stability of the fixed points in the MFT
and the  BBGKY are also the same. The fixed points in this case are
stable focus fixed point. The system go to the fixed points wavily.

When the parameters fall in the  range of
\begin{eqnarray}
N{g^2} < 64{\Gamma ^2} < N{g^2}\sqrt {1{\rm{ + }}3\left( {1{\rm{ + }}4/N} \right)},
\label{middle}
\end{eqnarray}
\begin{figure}[h]
\centering
\includegraphics[angle=0,width=0.45\textwidth]{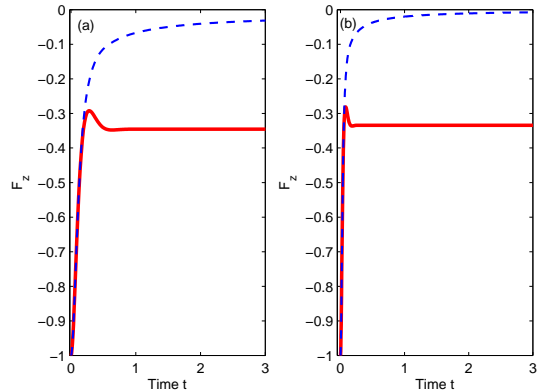}
\caption{(Color online) Red-thick line shows how the system goes to
the stable focus fixed points by MFT, while  blue-dashed line shows
how the system goes to the stable junction fixed point by  the BBGKY
in Eq.~(\ref{Beyondequation20}). Parameters chosen satisfying the
Eq.~(\ref{middle}) are $g=1, N=80$, $\Gamma=1.1192$, $\left| {{\psi
_0}} \right\rangle  = \left| {0} \right\rangle $ for (a), and
$\Gamma=3.9568, N=1000$, $\left| {{\psi _0}} \right\rangle = \left|
{0} \right\rangle $ for (b).} \label{oscillation:}
\end{figure}
the system in the BBGKY theory  Eq.~(\ref{Beyondequation20}) would
go to a stable junction fixed point , but by the MFT, the system
would approach to  a stable focus fixed point. We plot the time
evolution of ${F_z}$ in Fig.~\ref{oscillation:}. From the figure, we
can see that  the population difference ${F_z}$ in BBGKY theory
increases monotonously as $t$ increases (the blue-dashed line), but
it increases first  then  decreases   and finally reaches the stable
state in the MFT. In addition, comparing Fig.~(\ref{oscillation:})
(a) and (b), we can learn that in (a) $F_z$  changes slowly, while
in (b) it is faster, this is due to the difference of the dephasing
rate $\Gamma$.

\textbf{Before concluding the paper, we present a discussion on the
time-dependent many-body theory \cite{Javanainen2002,Naidon2008} and
the master equation approach in the two-mode approximation. We start
with the many-body description for the photoassociation in a uniform
Bose-Einstein condensate \cite{Naidon2008}. In the two-body case,
the system model  reduces to a set of coupled modes, two of them are
atoms in condensate and molecules. The other modes represent the
noncondensate atom pairs. This treatment is very similar to the
master equation description, when the noncondensate atom pairs are
treated as an environment. Then the elimination of the modes of
noncondensate atom pairs in the two-body theory would lead to
equations of motion (almost) equivalent to that in the master
equation description.}

\textbf{To be specific, we take the photoassociation of a
Bose-Einstein condensate \cite{Javanainen2002} as an example.  The
equation of motion of the system reads,}
\begin{eqnarray}
\dot{\alpha} &=& i\frac{\Omega}{\sqrt{2}}\,\alpha^*\beta,\nonumber\\
\dot{\beta}  &=& i\delta\beta +
i\frac{\Omega^*}{\sqrt{2}}\,\alpha^2+i\int
d\epsilon\,\xi(\epsilon)\,c_\epsilon,\nonumber\\
\dot{c}_\epsilon &=& -i\epsilon\,c_\epsilon + i
\xi^*(\epsilon)\beta\, \label{ja2002}
\end{eqnarray}
\textbf{where $\alpha=a/\sqrt{N}$, $\beta=\sqrt{2/N}b$, and
$c_{\epsilon}$ represent the c-number atomic, molecular and
non-condensate atom pair amplitudes, respectively. Formally
integrating the third equation of Eq. (\ref{ja2002}) and
substituting it into the second, with the Wigner-Weisskopf
approximation we have,}
\begin{eqnarray}
\dot{\alpha} &=& i\frac{\Omega}{\sqrt{2}}\,\alpha^*\beta,\nonumber\\
\dot{\beta}  &=& i\delta\beta + i\frac{\Omega^*}{\sqrt{2}}\,\alpha^2
-\Gamma \beta, \label{ja20021}
\end{eqnarray}
\textbf{where $\Gamma=\pi|\xi(0)|^2.$ On the other hand, under the
mean-field approximation, the coupled equations of $\alpha$ and
$\beta$ can be derived from a master equation with a dissipation
part,
$$\frac{\Gamma}{2}(2b\rho b^{\dagger}-\rho b^{\dagger}b-b^{\dagger}b\rho).$$
Although the descriptions based on the master equation and the
many-body theory yield a very similar equation of motion for the
condensed atoms and molecules in the photoassociation, the master
equation loses  (almost all) information of the non-condensate
atoms, as it is traced out as an environment. The benefit we gain
from the master equation description  is that it reduces the
calculation complexity. Nevertheless,  eliminating the environmental
degree of freedoms in the many-body theory in the mean-field
approximation can not give a mixed state for the reduced system.}

\section{Conclusion}
In this paper,  the dynamics of the atom-molecule conversion system
subject to dephasing noises has been explored. We find that the
fixed points given by the mean-field theory (MFT) and by numerically
solving the master equation are different, this indicates that the
mean-field theory is not a good treatment \textbf{at a long time scale} for the
atom-molecule conversion system. We further develop the  BBGKY
hierarchy truncation approach to study the atom-molecule conversion
system, fixed points are calculated and the stability around the
fixed points are discussed. We observe that  for a wide range of
parameters the stability around the fixed points are  the same in
the MFT and the BBGKY hierarchy truncation approach. The dynamics of
the atom-molecule conversion system is also explored, the results
suggest that the second-order of BBGKY hierarchy is a good approach
for the atom-molecule conversion system.
\ \ \\
This work is supported by the NSF of China under Grants Nos
61078011,  10935010 and 11175032.

\end{document}